\documentclass[sigconf]{acmart}

\AtBeginDocument{%
  \providecommand\BibTeX{{%
    \normalfont B\kern-0.5em{\scshape i\kern-0.25em b}\kern-0.8em\TeX}}}

\AtBeginDocument{%
 \providecommand\BibTeX{{%
    \normalfont B\kern-0.5em{\scshape i\kern-0.25em b}\kern-0.8em\TeX}}}
    
\usepackage{enumitem}
\usepackage{multirow}
\usepackage{makecell}
\usepackage{amsthm}
\usepackage{amsfonts}
\usepackage{url}
\usepackage{amsmath}
\usepackage{caption}
\usepackage{subcaption}

\newcommand{\ie}{\emph{i.e., }}

\newcommand{\etal}{\emph{et al. }}

\newcommand{\etc}{\emph{etc.}}
\newcommand{\wrt}{\emph{w.r.t. }}

\newcommand{\aka}{\emph{aka. }}

\newcommand{\wyw}[1]{{\color{black}{#1}}}

\acmConference[WSDM '24]{March 4--8, 2023}{Mérida, México}
\begin{document}
\settopmatter{printacmref=false}
% \fancyhead{}

%\title{Hierarchical Contrastive Transformer for Multimodal Bundle Construction}
%\title{Incorporating User Feedback to Multimodal Bundle Construction}
%\title{Leveraging Both User Feedback and Multimodal Features for Bundle Construction}
\title{Leveraging Multimodal Features and Item-level User Feedback for Bundle Construction}

\author{Yunshan Ma}
 \affiliation{
     \institution{National University of Singapore}
     \country{}
 }
\email{yunshan.ma@u.nus.edu}

\author{Xiaohao Liu}
 \affiliation{
     \institution{University of Chinese Academy of Sciences}
     \country{}
 }
\email{xiaohao.liu@hotmail.com} 

\author{Yinwei Wei}
\affiliation{
    \institution{Monash University}
    \country{}
}
\email{weiyinwei@hotmail.com}

\author{Zhulin Tao}
\authornote{Corresponding author.}
\affiliation{
    \institution{Communication University of China}
    \country{}
}
\email{taozhulin@gmail.com}

\author{Xiang Wang}
\authornote{Xiang Wang is also affiliated with Institute of Artificial Intelligence, Institute of Dataspace, Hefei Comprehensive National Science Center.}
\affiliation{
    \institution{University of Science and Technology of China}
    \country{}
}
\email{xiangwang1223@gmail.com}

\author{Tat-Seng Chua}
\affiliation{
    \institution{National University of Singapore}
    \country{}
}
\email{dcscts@nus.edu.sg}

\begin{abstract}
Automatic bundle construction is a crucial prerequisite step in various bundle-aware online services. Previous approaches are mostly designed to model the bundling strategy of existing bundles. However, it is hard to acquire large-scale well-curated bundle dataset, especially for those platforms that have not offered bundle services before. Even for platforms with mature bundle services, there are still many items that are included in few or even zero bundles, which give rise to sparsity and cold-start challenges in the bundle construction models. To tackle these issues, we target at leveraging  multimodal features, item-level user feedback signals, and the bundle composition information, to achieve a comprehensive formulation of bundle construction. Nevertheless, such formulation poses two new technical challenges: 1) how to learn effective representations by optimally unifying multiple features, and 2) how to address the problems of modality missing, noise, and sparsity problems induced by the incomplete query bundles. In this work, to address these technical challenges, we propose a \textbf{C}ontrastive \textbf{L}earning-enhanced \textbf{H}ierarchical \textbf{E}ncoder method (CLHE). Specifically, we use self-attention modules to combine the multimodal and multi-item features, and then leverage both item- and bundle-level contrastive learning to enhance the representation learning, thus to counter the modality missing, noise, and sparsity problems. Extensive experiments on four datasets in two application domains demonstrate that our method outperforms a list of SOTA methods. The code and dataset are available at \url{https://github.com/Xiaohao-Liu/CLHE}.

%\url{ https://anonymous.4open.science/r/CrossCBR-71B1/}.
\end{abstract}

% \ccsdesc[500]{Information systems~Recommender systems}

\maketitle

\section{Introduction} \label{sec:introduction}

Product bundling has been a popular and effective marketing strategy, tracing back from ancient commercial times and persisting through to the rapidly growing e-commerce and online services today. By combining a set of individual items into a bundle, both the sellers (or service providers) and consumers can benefit a lot from multiple aspects, including the reduced cost of packaging, shipment, and installation, to promoting sales of old or new items by combining them with some popular or essential items with discounts. To implement product bundling, the first and foremost step is constructing bundles from individual items, \aka bundle construction, which is traditionally carried out by human experts. However, the explosive growth of item sets poses significant challenges to such high-cost manual approaches. Hence, automatic approaches to bundle construction are imperative and have garnered more and more attention in recent years.

\begin{figure}
    \centering
    \includegraphics[width = 0.99\linewidth]{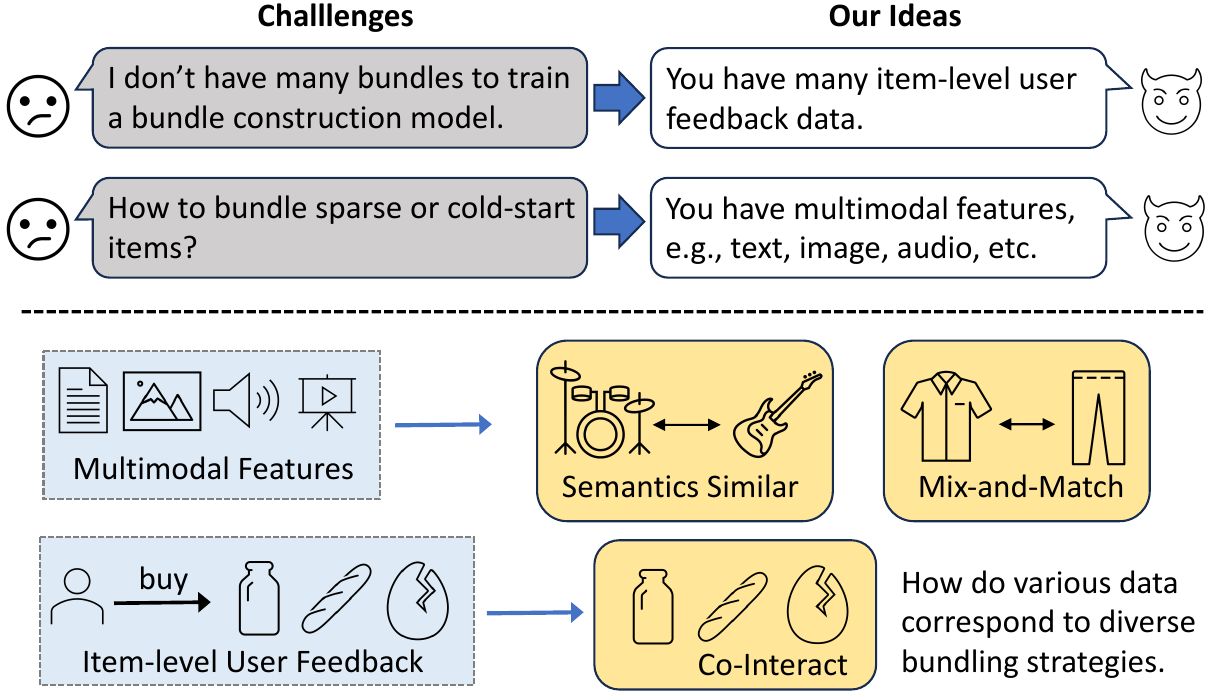}
    \vspace{-0.1in}
    \caption{The motivations of leveraging multimodal features and item-level user feedback for bundle construction.}
    \label{fig:motivation}
    \vspace{-0.2in}
\end{figure}

\wyw{By analyzing prior studies, we find that they mostly build the bundles based on the co-occurrence relationship of items in existing training bundles. However, there are two key} problems that have not been well studied: 1) previous approaches heavily rely on large-scale high-quality bundle dataset for training, and 2) they cannot properly handle the sparsity and cold-start issues. First, most previous bundle construction methods require high-quality supervision signals from a large set of well-curated bundles. However, there is a dilemma in such an approach especially for platforms that have not offered bundle service before or have just deployed bundle service for a short period of time, it is difficult for such platforms to collect sufficient bundle data for training. Second, even for platforms with mature bundle services, the situation is far from ideal due to the various cold-start problems. On the one hand, there are quite a number of items that are only involved in a few bundles, consequently, it is challenging to obtain informative representations for these sparse items to construct new bundles. Worse still, there many new items, which haven't been part of previous bundles while are continuously pushed online, and how to swiftly bundle these cold-start items with existing \textit{warm} items is crucial for platforms to promote new products and keep sustained growth. 

Addressing these challenges, instead of seeking any silver bullet, we are more keen on practical solutions that make full use of the large amount of easy-to-access resources: multimodal features and item-level user feedback. 
\wyw{The motivation behind this solution is that these data are well aligned to diverse bundling strategies. }
First, multimodal features, such as text, image, and audio, contain rich semantic information that is helpful to find either similar or compatible items and form bundles, as shown in Figure~\ref{fig:motivation}. More importantly, most items, even those sparse and newly introduced items, usually have one or multiple such features. A plethora of previous efforts, such as personalized recommendation~\cite{MMGCN2019}, have demonstrated the efficacy of multimodal features in handling sparse and cold-start items. Second, item-level user feedback information endows precious crowd-sourcing knowledge that is crucial to bundle construction. Intuitively, the items that users frequently co-interact with are strong candidates for bundling. More importantly, a large amount of such user feedback signals are available even to platforms that do not offer bundle services. Compared with previous works~\cite{MMOCM2021}, we pioneer the integration of multimodal features and item-level user feedback for bundle construction. 

Given the outlined motivations, we aim to leverage both multimodal features and item-level user feedback, along with the existing bundles, to develop a comprehensive model for bundle construction. However, it is non-trivial to design a model to capture all three types of information and achieve optimal bundle construction performance. First, how to learn effective representations in each modality and well capture the cooperative association among the three modalities is a key challenge. Second, some items might not be associated with user feedback or affiliated to bundles comprehensively, thus the so-called modality-missing issue may degrade the modeling capability. What's more, during the inference stage of bundle construction, we usually need to provide several seed items as a partial bundle to initiate the construction process. However, the incompleteness of the partial bundle imposes noise and sparsity challenges to the bundle representation learning, which will impede the bundle construction performance.

In this work, to address the aforementioned challenges, we propose a \textbf{C}ontrastive \textbf{L}earning-enhanced \textbf{H}ierarchical \textbf{E}ncoder (CLHE) for bundle construction. In order to obtain the representations of items, we make use of the recently proposed large-scale multimodal foundation models (\ie BLIP~\cite{BLIP2022} and CLAP~\cite{CLAP2022}) to extract the multimodal features of items. Concurrently, we pre-train a collaborative filtering (CF)-based model (\ie LightGCN~\cite{LightGCN2020}) to obtain the items' representations that preserve the user feedback information. Then, we employ a hierarchical encoder to learn the bundle representation, where the self-attention mechanism is devised to expertly fuse multimodal information and multi-item representations. To tackle the modality missing problem and the sparsity/noise issues induced by the incomplete partial bundle, we employ two levels of contrastive learning~\cite{SGL2021,CrossCBR2022}, \ie item-level and bundle-level, to fully take advantage of the self-supervision signals. We conduct experiments on four datasets from two domains, and the results demonstrate that our method outperforms multiple leading methods. Various ablation and modal studies further justify the effectiveness of key modules and demonstrate multiple crucial properties of our proposed model. We summarize the key contributions of this work as follows:
\begin{itemize}[leftmargin=*]
    \item We introduce a pioneering approach to bundle construction by holistically combining multimodal features, item-level user feedback, and existing bundles. This integration addresses prevailing challenges such as data insufficiency and the cold-start problem.
    \item We highlight multiple technical challenges of this new formulation and propose a novel method of CLHE to tackle them.
    \item Our method outperforms various leading methods on four datasets from two application domains with different settings, and further diverse studies demonstrate various merits of our method. 
\end{itemize}

%\begin{figure}
%    \centering
%    \includegraphics[width = 0.95\linewidth]{figures/motivation.pdf}
%    %\vspace{-5pt}
%    \caption{The main motivation of context-aware event forecasting. (a) Most current events fall in the coarse-grained and higher level types of the ontology, while more informative fine-grained events are fewer. (b) Out-of-ontology and diverse contexts affect events. Context can provide more fine-grained information and enhance the event forecasting performance.}
%    \label{fig:motivation}
%    \vspace{-15pt}
%\end{figure}
\section{Methodology} \label{sec:methodology}
\begin{figure*}
    \centering
    \includegraphics[width=0.99\textwidth]{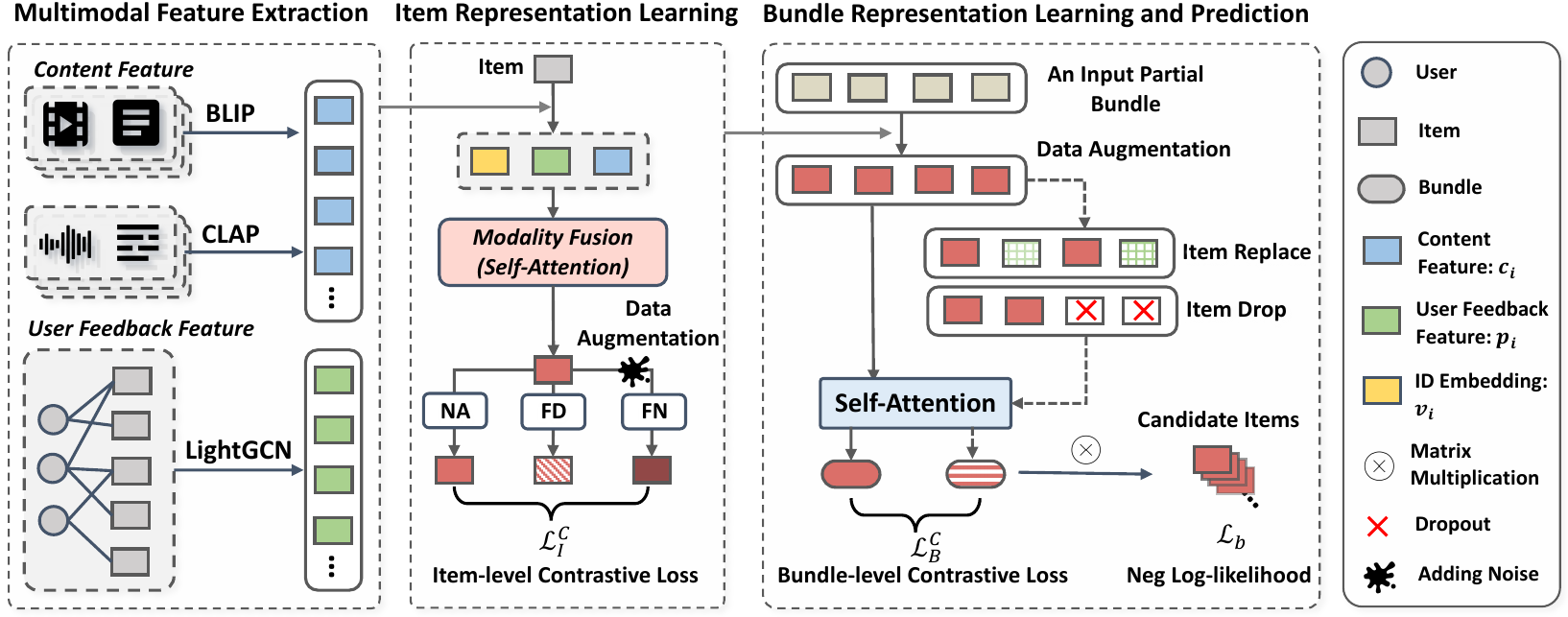}
    \vspace{-0.1in}
    \caption{The overall framework of our proposed method CLHE, which consists of two main components: hierarchical encoder (\aka multimodal feature extraction, item and bundle representation learning) and contrastive learning.}
    \vspace{-0.1in}
    \label{fig:framework}
\end{figure*}

We first formally define the problem of bundle construction by considering all three types of data. Then we describe the details of our proposed method CLHE (as shown in Figure~\ref{fig:framework}).

%including the hierarchical encoder for multimodal bundle representation learning, the contrastive learning modules, and the prediction and optimization.

\subsection{Problem Formulation} \label{subsec:prob_form}
 % the item set and user-item interactions are univeral, while the bundles should be classified into training set and testing set. Just like image classification problem.

Given a set of items $\mathcal{I}=\{i_1, i_2, \cdots, i_N\}$, each item has a textual input $t_i$, which can be its title, description, or metadata, and a media input $m_i$, which can be an image, audio, or video of the item. In addition, for the items that have been online for a while, we have collected some item-level user feedback data, which is denoted as a user-item interaction matrix $\mathbf{X}_{M \times N}=\{x_{ui}|u\in{\mathcal{U}},i\in{\mathcal{I}}\}$, where $\mathcal{U}=\{u_1, u_2, \cdots, u_M\}$ is the user set. We define a bundle as a set of items, denoted as $b=\{i_1, i_2, \cdots, i_n\}$, where $n=\lvert b \rvert$ is the size of the bundle. Given a partial bundle $b_s \subset b$ (\ie a set of seed items), where $\lvert b_s \rvert < \lvert b \rvert$, the bundle construction model targets at predict the missing items $i \in b \setminus b_s$. 
We have a set of known bundles for training, denoted as $\mathcal{B}=\{b_1, b_2, \cdots, b_O\}$ and a set of unseen bundles for testing, denoted as $\mathcal{\bar{B}}=\{b_{O+1}, b_{O+2}, \cdots, b_{O+\bar{O}}\}$, where $O$ is the number of training bundles and $\bar{O}$ is the number of testing bundles. We would like to train a model based on the training set $\mathcal{B}$, for an unseen bundle $b \in \mathcal{\bar{B}}$, when given a few seed items $\bar{b}_s$, \aka the partial bundle, the model can predict the missing items $b \setminus b_s$ thus to construct the entire bundle.

\subsection{Hierarchical Encoder}
%We employ a hierarchical encoder to learn the multimodal bundle representation. Specifically, we first extract the multimodal features using the cutting-edge multimodal foundation models, and in parallel, we pre-train a CF-based model to obtain the item-level user feedback feature. Then, we devise the first self-attention encoder to learn the item representation by fusing the multimodal features. Following on, we employ the second self-attention encoder to aggregate multiple items' representations within the bundle, thus yielding the entire bundle representation. 
We utilize a hierarchical encoder for multimodal bundle representation. Initially, we extract multimodal features using multimodal foundation models, while concurrently pre-training a CF-based model to capture item-level user feedback. Subsequently, a self-attention encoder is introduced to integrate these multimodal features, resulting in a fused item representation. Another self-attention encoder then aggregates these representations, producing a comprehensive bundle representation.

\subsubsection{Item Representation Learning}
We first detail the feature extraction process and then present the self-attention encoder.

\textbf{Multimodal Feature Extraction}. We seek large-scale multimodal foundation models to extract the textual and media features of items. Compared with previous uni-modal feature extractors, such as in Computer Vision (CV)~\cite{RESNET2016,SwinTransformer2021}, Natural Language Processing (NLP)~\cite{BERT2019,RoBERTa2019}, or audio~\cite{PANNS2020,HTSAT2022}, multimodal foundation models are more powerful to capture the multimodal semantics of the input data, which have demonstrated to be effective in transferring or generalizing to various downstream tasks. Concretely, for image data, we use the BLIP~\cite{BLIP2022} model to extract both textual and visual features. For audio data, we use the CLAP~\cite{CLAP2022} model to extract textual and audio features. After the feature extraction, we obtain the textual feature $\mathbf{t}_i \in \mathbb{R}^{768}$ and media feature $\textbf{m}_i \in \mathbb{R}^{768}$. Given their shared representation space, we perform a simple average pooling over them, resulting in the content feature of the item, denoted as $\mathbf{c}_i= \text{average}(\mathbf{t}_i, \mathbf{m}_i)$.
%, where $\text{average}(\cdot)$ is the average pooling function.

\textbf{Item-level User Feedback Feature Extraction}. We employ the well-performing CF-based model, \ie LightGCN~\cite{LightGCN2020}, to obtain item representations from user feedback. Specifically, we devise a bipartite graph based on the user-item interaction matrix, then train a LightGCN~\footnote{Other CF-based models can also be used.} model over the bipartite graph, denoted as:
\begin{equation} \label{eq_1}
\left\{
\begin{aligned}
    \mathbf{p}_{u}^{(k)} &= \sum_{i \in \mathcal{N}_u}{\frac{1}{\sqrt{|\mathcal{N}_u|}\sqrt{|\mathcal
    {N}_i|}}\mathbf{p}^{(k-1)}_i}, \\
    \mathbf{p}_{i}^{(k)} &= \sum_{u \in \mathcal{N}_i}{\frac{1}{\sqrt{|\mathcal{N}_i|}\sqrt{|\mathcal{N}_u|}}\mathbf{p}^{(k-1)}_{u}},
\end{aligned}
\right.
\end{equation}
where $\mathbf{p}_{u}^{(k)}, \mathbf{p}_{i}^{(k)} \in \mathbb{R}^{d}$ are embeddings for user $u$ and item $i$ at the $k$-th layer, and $d$ is the dimensionality of the hidden representation; $\mathcal{N}_u$ and $\mathcal{N}_i$ are the neighbors of the user $u$ and item $i$ in the user-item interaction graph. We only make use of the item representation $\mathbf{p}_{i}$, which captures the item-level user feedback information. It is tailored by aggregating the item representations over $K$ layers' propagation, denoted as: 
\begin{equation} \label{eq_2}
    \mathbf{p}_{i} = \frac{1}{K} \sum^K_{k=0}{\mathbf{p}^{(k)}_{i}}.
\end{equation}

\textbf{ID Embedding Initialization}. We also initialize an id embedding $\mathbf{v}_i \in \mathbb{R}^d$ for each item to capture its bundle-item affiliation patterns. Please note that for those items (both during training and testing) that do not have user feedback features, we copy the content feature to its user feedback feature slot. Analogously, for the cold-start item that do not have an id embedding, we copy its corresponding content feature to take the slot.

\textbf{Modality Fusion via Self-attention}. Given the three types of features, \ie $\mathbf{c}_i$, $\mathbf{p}_i$, and $\mathbf{v}_i$, we first apply a feature transformation layer to project the multimodal and user-feedback features into the same latent space with the id embeddings, then we concatenate all the three features into a feature matrix $\mathbf{F}_{i} \in \mathbb{R}^{3 \times d}$, denoted as:
\begin{equation} \label{eq_3}
    \mathbf{F}_{i} = \text{concat}(\mathbf{c}_i\mathbf{W}_c, \mathbf{p}_i\mathbf{W}_p, \mathbf{v}_i),
\end{equation}
where $\mathbf{W}_c \in \mathbb{R}^{768 \times d}$ and $\mathbf{W}_p \in \mathbb{R}^{768 \times d}$ are the transformation matrices for multimodal and user-feedback features, respectively; $\text{concat}(\cdot)$ is the concatenation function. Then, we devise a self-attention layer to model the correlations of multiple features, denoted as:
\begin{equation} \label{eq_4}
\left\{
\begin{aligned}
    \mathbf{A}_i^{(l)} &= \frac{1}{\sqrt{d}}\mathbf{\hat{F}}_i^{(l-1)}\mathbf{W}_I^K \bigl(\mathbf{\hat{F}}_i^{(l-1)}\mathbf{W}_I^Q \bigr)^{\intercal}, \\
    \mathbf{\Tilde{F}}_i^{(l)} &= \text{softmax} \bigl(\mathbf{A}_i^{(l)} \bigr)\mathbf{\hat{F}}_i^{l-1},
\end{aligned}
\right.
\end{equation}
where $\mathbf{W}_I^K \in \mathbb{R}^{d \times d}$ and $\mathbf{W}_I^Q \in \mathbb{R}^{d \times d}$ are the trainable parameters for this item-level encoder to project the input feature embeddings into the key and value spaces; $\mathbf{\hat{F}}_i^{(l)} \in \mathbb{R}^{3 \times d}$ is the hidden feature representations in the intermediate layer $l$, and $\mathbf{\hat{F}}_i^{(0)}=\mathbf{F}_i$; $\text{softmax}(\cdot)$ is the softmax function and $\mathbf{\Tilde{F}}_i^{L}$ denotes the features' representations after $L$ layers of self-attention. We then average the multiple features to obtain the item representation $\mathbf{f}_i \in \mathbb{R}^{d}$ after multimodal fusion, formally defined as:
\begin{equation} \label{eq_5}
    \mathbf{f}_{i} = \text{average} \bigl({\mathbf{\Tilde{F}}_i^{(L)}} \bigr).
\end{equation}

\subsubsection{Bundle Representation Learning}
After obtaining the item representation, we build a second self-attention module to learn the representation of the given partial bundle. For a certain partial bundle $b_s$, its representation $\mathbf{e}_{b_s}$ \footnote{For simplicity, we omit the subscript $s$ in $b_s$ and just use $b$ and $\mathbf{e}_{b}$ to represent the partial bundle and its representation if there is no ambiguity.} is learned by:
\begin{equation} \label{eq_6}
\left\{
\begin{aligned}
    \mathbf{A}_b^{(z)} &= \frac{1}{\sqrt{d}}\mathbf{\hat{E}}_b^{(z-1)}\mathbf{W}_B^K \bigl(\mathbf{\hat{E}}_b^{(z-1)}\mathbf{W}_B^Q \bigr)^{\intercal}, \\
    \mathbf{\Tilde{E}}_b^{(z)} &= \text{softmax} \bigl(\mathbf{A}_b^{(z)} \bigr)\mathbf{\hat{E}}_b^{z-1},
\end{aligned}
\right.
\end{equation}
where $\mathbf{W}_B^K \in \mathbb{R}^{d \times d}$ and $\mathbf{W}_B^Q \in \mathbb{R}^{d \times d}$ are the trainable parameters in the bundle-level to project the input item embeddings into the key and value spaces; $\mathbf{\hat{E}}_b^{(z)} \in \mathbb{R}^{\lvert b \rvert \times d}$ is the hidden representations in the middle layer $z$, and $\mathbf{\hat{E}}_b^{(0)}=\text{concat}(\{\mathbf{f}_i\}_{i \in b})$; $\mathbf{\Tilde{E}}_b^{Z}$ denotes the features' representations after $Z$ layers of self-attention. We then average the multiple features to obtain the item representation $\mathbf{e}_b$ after multimodal fusion, formally defined as:
\begin{equation} \label{eq_7}
    \mathbf{e}_{b} = \text{average} \bigl({\mathbf{\Tilde{E}}_b^{(Z)}} \bigr).
\end{equation}

\begin{table*}[t]
\begin{center}
\vspace{-0.1in}
\caption{The statistics of the four datasets on two different domains.}
\label{tab:dataset}
\vspace{-0.1in}
\resizebox{0.85\textwidth}{!}{
    \begin{tabular}{l ccc cc cccc}
        \toprule
        Dataset & \#U & \#I & \#B & \#B-I & \#U-I & \#Avg.I/B & \#Avg.B/I & \#Avg.I/U & \#Avg.U/I \\
        \midrule
        POG             & 17,449    & 48,676  & 20,000 & 72,224    & 237,519    & 3.61  & 1.48 & 13.61  & 4.88   \\  
        POG\_dense      & 2,311,431 & 31,217  & 29,686 & 105,775   & 6,345,137  & 3.56  & 3.39 & 2.75   & 203.26 \\ 
        Spotify         & 118,994   & 254,155 & 20,000 & 1,268,716 & 36,244,806 & 63.44 & 4.99 & 304.59 & 142.61 \\
        Spotify\_sparse & 118,899   & 213,325 & 12,486 & 549,900   & 32,890,315 & 44.04 & 2.58 & 276.62 & 154.18 \\
        \bottomrule
    \end{tabular}
}
\end{center}
\vspace{-0.1in}
\end{table*}

\subsection{Contrastive Learning}
Even though the hierarchical encoder can well attain the correlations among multiple features and multiple items, it still suffers from noise, sparsity, or even cold-start problems in both item and bundle levels. Specifically, at the item level, the items that have fewer user feedbacks or are involved in fewer bundles during training may also be prone to deteriorate representations, which is the so-called sparsity issue. Even worse, some cold-start items may have never interacted with any users or been included in any bundles before, therefore, the cold-start problem will severely deteriorate the representation quality. Second, at the bundle level, the partial bundle's representation is susceptible to noise and sparsity issues. Instead of a complete bundle that is sufficient to depict all the functionalities or properties of the bundle, the given partial bundle only encompasses some of the items. Consequently, the bundle representation may be biased due to the arbitrary seed items. 

To tackle these problems, we aim to harness contrastive learning over both item and bundle levels to mine the self-supervision signals. Recently, contrastive learning has achieved great success in various tasks, including CV~\cite{SimCLR2020}, NLP~\cite{SimCSE2021}, and recommender systems~\cite{SGL2021}. The main idea is to first corrupt the original data and generate some augmented views for the same data point, and then leverage an InfoNCE loss to pull close the representations across multiple augmented views for the same data point, while pushing away the representations of different data points. Therefore, the representations could be more robust to combat noise and sparsity. 

\subsubsection{Item-level Contrastive Learning} For each item $i$, we tailor its representation $\mathbf{f}_i$ in Equation~\ref{eq_5}. We leverage various data augmentations to generate the augmented view $\mathbf{f}_i^{\prime}$. The item-level data augmentation methods we used include: 1) No Augmentation (NA)~\cite{CrossCBR2022}: just use the original representation as the augmented feature without any augmentation; 2) Feature Noise (FN)~\cite{xSimGCL2022}: add a small-scaled random noise vector to the item’s features; 3) Feature Dropout (FD)~\cite{SGL2021}: randomly dropout some values over the feature vectors; and 4) Modality Dropout (MD): dropout the whole feature of a randomly selected modality on a randomly selected item. Then, we use the InfoNCE~\cite{CrossCBR2022} to generate the item-level contrastive loss, denoted as:
\begin{equation} \label{eq_8}
    \mathcal{L}^C_{I} = \frac{1}{|\mathcal{I}|}\sum_{i \in \mathcal{I}}{-\text{log}\frac
            {\text{exp}({\text{cos}(\mathbf{f}_{i}, \mathbf{f}_{i}^{\prime})/\tau})}
            {\sum_{v \in \mathcal{I}}{\text{exp}({\text{cos}(\mathbf{f}_{i}, \mathbf{f}_{v}^{\prime})/\tau})}}},
\end{equation}
where $\text{cos}(\cdot)$ is the cosine similarity, and $\tau$ is the temperature. 

\subsubsection{Bundle-level Contrastive Learning}
For each bundle $b$ and its original representation $\mathbf{e}_b$, we also implement various data augmentations to generate an augmented view $\mathbf{e}_b^{\prime}$. The data augmentation methods we leveraged include: 1) Item Dropout (ID): randomly dropout some items in the bundle; and 2) Item Replacement (IR): randomly select some items in the bundle and replace them with some other items that have not appear in the bundle. Following on, the bundle-level contrastive loss is tailored by:
\begin{equation} \label{eq_9}
    \mathcal{L}^C_{B} = \frac{1}{|\mathcal{B}|}\sum_{b \in \mathcal{B}}{-\text{log}\frac
            {\text{exp}({\text{cos}(\mathbf{e}_{b}, \mathbf{e}_{b}^{\prime})/\tau})}
            {\sum_{v \in \mathcal{B}}{\text{exp}({\text{cos}(\mathbf{e}_{b}, \mathbf{e}_{v}^{\prime})/\tau})}}}.
\end{equation} 

\begin{table*}[t]
\caption{The overall performance of our CLHE and baselines. "\%Improv." denotes the relative improvement over the strongest baseline. The best baselines are underlined.}
\vspace{-0.1in}
\label{tab:overall_performance_mm}
\centering
\setlength{\tabcolsep}{1mm}{
    \resizebox{0.85\textwidth}{!}{
        \begin{tabular}{l cc cc cc cc}
        \toprule
         \multirow{2.4}{*}{Models} & \multicolumn{2}{c}{POG} & \multicolumn{2}{c}{POG\_dense} & \multicolumn{2}{c}{Spotify} & \multicolumn{2}{c}{Spotify\_sparse} \\
         \cmidrule(lr){2-3} \cmidrule(lr){4-5} \cmidrule(lr){6-7} \cmidrule(lr){8-9} & Rec@20 & NDCG@20 & Rec@20 & NDCG@20 & Rec@20 & NDCG@20 & Rec@20 & NDCG@20 \\
        \midrule
         \textbf{MultiDAE}             & 0.0119 & 0.0063 & 0.3213 & 0.2179 & 0.0578 & 0.0944 & 0.0506 & 0.0574  \\
         \textbf{MultiVAE}            & 0.0196 & 0.0104 & 0.3221 & 0.2086 & 0.0400 & 0.0656 & 0.0325 & 0.0365  \\
         \textbf{Bi-LSTM}        & 0.0170 & 0.0097 & 0.2932 & 0.1745 & 0.0833 & 0.1486 & 0.0645 & 0.0822  \\
         \textbf{Hypergraph}     & 0.0207 & 0.0111 & 0.3063 & 0.2256 & 0.0572 & 0.0941 & 0.0529 & 0.0590  \\
         \textbf{Tranformer}     & \underline{0.0215} & 0.0114 & \underline{0.3525} & \underline{0.2527} & 0.0875 & 0.1460 & 0.0768 & 0.0902  \\
         \textbf{TranformerCL}  & 0.0202 & \underline{0.0134} & 0.3170 & 0.2374 & \underline{0.1014} & \underline{0.1696} & \underline{0.0874} & \underline{0.1062}  \\
        \midrule
         \textbf{CLHE (ours)} & \textbf{0.0284} & \textbf{0.0193} & \textbf{0.3811} & \textbf{0.2773} & \textbf{0.1081} & \textbf{0.1806} & \textbf{0.0980} & \textbf{0.1212}  \\
         \textbf{\%Improv.}      & 32.45  & 44.03  & 8.13   & 9.71   & 6.61   & 6.49   & 12.12  & 14.15   \\
        \bottomrule
        \end{tabular}
    }
}
\vspace{-0.1in}
\end{table*}

\subsection{Prediction and Optimization}
After obtain the partial bundle representation $\mathbf{e}_{b_s}$ and the item representations $\mathbf{f}_i$, we leverage the inner-product function to induce the score $\hat{y}_{b_s,i}$ that indicates the possibility of item $i$ being included into bundle $b$ to make it complete, defined as:
\begin{equation} \label{eq_10}
    \hat{y}_{b_s,i} = \mathbf{e}_{b_s} \mathbf{f}_i^{\intercal}.
\end{equation} 

To optimize our model, we follow the previous approaches~\cite{MultiDAE2016,MultiVAE2018} and leverage the negative log-likelihood loss, therefore, the loss for bundle $b$ is denoted as:
% \begin{equation} \label{eq_11}
%     \mathcal{L}_b = \sum_{i \in b \setminus b_s}{y_{b_s,i}\text{log}\sigma(\hat{y}_{b_s,i})+(1-y_{b_s,i})\text{log}(1-\sigma(\hat{y}(b_s,i)))},
% \end{equation}
% where $\sigma(x)=1/(1+\text{exp}(-x))$ is the logistic function. 
\begin{equation} \label{eq_11}
    \mathcal{L}_b = \frac{1}{|\mathcal{I}|}\sum_{i \in \mathcal{I}}{- y_{b_s,i}\text{log}\pi_{b_s}(\hat{y}_{b_s,i})},
\end{equation}
where $\pi(\cdot)$ is the softmax function which produces the probabilities over the entire items.
In collaboration with the contrastive loss and regularization, we have the final loss, denoted as:
\begin{equation} \label{eq_12}
    \mathcal{L} = \frac{1}{\lvert \mathcal{B} \rvert} \sum_{b \in \mathcal{B}}{\mathcal{L}_b} + \alpha_1\mathcal{L}_I^C + \alpha_2\mathcal{L}_B^C + \beta{\Vert \mathbf{\Theta} \rVert}_2^2,
\end{equation}
where ${\alpha}_1$, ${\alpha}_2$ and $\beta$ are hyper-parameters to balance different loss terms, ${\Vert \mathbf{\Theta} \rVert}_2^2$ is the L2 regularization term, and ${\Vert \mathbf{\Theta} \rVert}$ denotes all the trainable parameters in our model.

%\subsection{Discussion}
%We would like to highlight that our method has strong capabilities in handling sparsity and cold-start issues for items. 

%First, for items that have few bundle-item connections, we cannot learn a good item representation only relying on the bundle-item affiliations. By incorporating the content and user feedback features into the modeling, these items would get more informative representations. 

%Second, we argue that the relational information of bundle-item and user-item will be complementary with each other. That is to say, if an item has few bundle-item connections while has rich user-item interactions, the user-item interactions would supplement and enhance the item representation. 
\section{Experiments} \label{sec:experiment}
We evaluate our proposed methods on two application domains of product bundling, \ie fashion outfit and music playlist. We are particularly interested in answering the research questions as follow:

\begin{itemize}[leftmargin=*]
    \item \textbf{RQ1: } Does the proposed CLHE method beat the leading methods?
    \item \textbf{RQ2: } Are the key modules, \ie hierarchical transformer and contrastive learning, effective?
    \item \textbf{RQ3: } How does out method work in countering the problems of cold-start items, modality missing, noise and sparsity of the partial bundle? How the detailed configurations affect its performance and how about the computation complexity?
\end{itemize}

\subsection{Experimental Settings}
There are various application scenarios that are suitable for product bundling, such as e-commerce, travel package, meal, \etc, each of which has one or multiple public datasets. However, only datasets that include all the multimodal item features, user feedback data, and bundle data can be used to evaluate our method. Therefore, we choose two representative domains, \ie fashion outfit and music playlist. We use the POG~\cite{POG2019} for fashion outfit. For the music playlist, we use the Spotify~\cite{SpotifyChallengeRecsys2018} dataset for the bundle-item affiliations, and we acquire the user feedback data from the Last.fm dataset~\cite{LastFM2011}. Since the average bundle size is quite small in POG (it makes sense for fashion outfit), we re-sample a second version POG\_dense which has denser user feedback connections for each item. In contrast, the average bundle size in Spotify dataset is large, thus we sample a sparser version Spotify\_sparse, which has smaller average bundle size. To be noted, we keep the integrity of all the bundles in all the versions, which means we do not corrupt any bundles during the sampling. For each dataset, we randomly split all the bundles into training/validation/testing set with the ratio of 8:1:1. The statistics of the datasets are shown in Table~\ref{tab:dataset}. We use the popular ranking protocols of Recall@K and NDCG@K as the evaluation metric, where K=20. 

\subsubsection{Compared Methods} Due to the new formulation of our work, there are no previous works that have exactly same setting with ours. Therefore, we pick several leading methods and adapt them to our settings. For fair comparison, all the baseline methods use all the three types of extracted features that are same with our method. In addition, they all use the same negative log-likelihood loss function.

\begin{itemize}[leftmargin=*]
    \item \textbf{MultiDAE}~\cite{MultiDAE2016} is an auto-encoder model which uses an average pooling to aggregate the items' representations to get the bundle representation. 
    
    \item \textbf{MultiVAE}~\cite{MultiVAE2018} is an variational auto-encoder model which employ the variational inference on top of the MultiDAE method.
    
    \item \textbf{Bi-LSTM}~\cite{Bi-LSTM2017} treats each bundle as a sequence and uses bi-directional LSTM to learn the bundle representation. 

    \item \textbf{Hypergraph}~\cite{Hypergraph2022} formulates each bundle a hyper-graph and devises a GCN model to learn the bundle representation.

    \item \textbf{Transformer}~\cite{BRUCE2022,BundleGT2023} tailors a transformer to capture the item interactions and generate the bundle representation.

    \item \textbf{TransformerCL} is the version that we add bundle-level contrastive loss to the above Transformer model.
\end{itemize}

\subsubsection{Hyper-parameter Settings}
The embedding and hidden representation size is 64, and we use Xavier~\cite{Xavier2010} initialization, batch size 2048, and Adam optimizer~\cite{Adam2014}.
We find the optimal hyper-parameter setting by adopting grid search. 
Wherein, learning rate is searched in range of $\{10^{-2}, 2 \times 10^{-2}, 10^{-3}, 2\times 10^{-3}, 10^{-4}, 2 \times 10^{-4}\}$ and $\beta$ is tuned in range of $\{10^{-3}, 10^{-4}, 10^{-5}, 10^{-6}\}$.
In most cases, the optimal value of learning rate is $10^{-3}$ and the one of $\beta$ is $10^{-5}$.
According to the contrastive learning, we search ${\alpha}_1,{\alpha}_2$ and $\tau$ in range of $\{0.1, 0.2, 0.5, 1, 2\}$ and $\{0.1, 0.2, 0.5, 1, 2, 5\}$, respectively.
Besides, we dropout features and modalities in augmentation step randomly with the ratio in range of $\{0, 0.1, 0.2, 0.5\}$ and add noise with a weight in range of $\{0.01, 0.02, 0.05, 0.1\}$. We search the number of propagation layers $K,L,Z$ from $\{1, 2, 3\}$.
For the baselines, we follow the designs in their articles to achieve the best performance. Certainly, we keep the same settings to ensure a fair comparison. 

% lr: {1e-2, 2e-2, 1e-3, 2e-3, 1e-4, 2e-4}, \beta: {1e-3, 1e-4, 1e-5, 1e-6}
% \alpha: {0.1, 0.2, 0.5, 1, 2}, \tau: {0.1, 0.2, 0.5, 1, 2, 5}

\subsection{Overall Performance Comparison (RQ1)} \label{subsec:overall_performance}
Table~\ref{tab:overall_performance_mm} shows the overall performance comparison between our model CLHE and the baseline methods. We have the following observations. First, our method beats all the baselines on all the datasets, demonstrating the competitive performance of our model. Second, over the baselines, Transformer and TransformerCL achieve the best performance, showing that the self-attention mechanism and contrastive learning can well preserve the correlations among items within the bundle, thus yielding good bundle representations. Third, comparing the results between different versions of dataset, we find that: 1) the performance on POG\_dense is much larger than that on POG due to denser user-item interactions, demonstrating that user feedback information is quite helpful to the performance; 2) the performance of Spotify\_sparse is relatively smaller than that on Spotify since the sparser bundle-item affiliation data, justifying our hypothesis that large-scale and high-quality bundle dataset is vital to bundle construction. Finally, we have an interesting observation that the performance improvements on the four datasets is negatively correlated with "\textit{\#Avr.B/I}", as shown in Table~\ref{tab:dataset}, in another word, in scenarios that items are included in fewer bundles (\ie the dataset include more sparse items), our method performs even better. This phenomenon further justifies the advantage of our method in countering the issue of sparse items.

\begin{table}[t]
\begin{center}
%\vspace{-0.1in}
\caption{Ablation study of the hierarchical encoder and contrastive learning (the performance is NDCG@20).}
\label{tab:ablation_study}
\vspace{-0.1in}
\resizebox{0.48\textwidth}{!}{
    \begin{tabular}{llcccc}
        \toprule
         \multicolumn{2}{l}{Settings} & POG & POG\_dense & Spotify & Spotify\_sparse \\
        \midrule
         \multicolumn{2}{l}{CLHE} & 0.0193 & 0.2773 & 0.1806 & 0.1212 \\
        \midrule
         \multicolumn{2}{l}{w/o user feedback} & 0.0168 & 0.2733 & 0.1695 & 0.1174 \\
        \midrule
        \multirow{3}{*}{SelfAtt.} 
         & w/o item   & 0.0168 & 0.2551 & 0.1334 & 0.0822 \\ 
         & w/o bundle & 0.0127 & 0.2141 & 0.1785 & 0.1210 \\
         & w/o both   & 0.0034 & 0.20647 & 0.0418 & 0.0334 \\
        \midrule
        \multirow{3}{*}{CL} 
         & w/o item   & 0.0171 & 0.2742 & 0.1735 & 0.1203 \\ 
         & w/o bundle & 0.0176 & 0.2598 & 0.1740 & 0.1123 \\
         & w/o both   & 0.0178 & 0.2662 & 0.1730 & 0.1084 \\
        \bottomrule
    \end{tabular}
}
\end{center}
\vspace{-0.1in}
\end{table}

\subsection{Ablation Study of Key Modules (RQ2)} \label{subsec:ablation_study}
To further evaluate the effectiveness of the key modules of our model, we conduct a list of ablation studies and the results are shown in Table~\ref{tab:ablation_study}.
First and foremost, we aim to justify the effectiveness of the user feedback features. Thereby, we remove the user feedback features from our model (\ie remove $\mathbf{p}_i$ from $\mathbf{f}_i$) and build an ablated version of model, \ie \textit{w/o user feedback}. According to the result in Table~\ref{tab:ablation_study}, after removing user feedback features, the performance reduces clearly, verifying that user feedback feature is significant for bundle construction.
Second, we would like to evaluate whether each component of the hierarchical encoder is useful. We progressively remove the two self-attention modules from our model and replace them with an vanilla average pooling, thus yielding three ablated models, \ie \textit{w/o item}, \textit{w/o bundle}, and \textit{w/o both}. The results in Table~\ref{tab:ablation_study} show that the removal of either self-attention modules causes performance drop. These results further verify the efficacy of our self-attention-based hierarchical encoder framework. 
Third, to justify the contribution of contrastive learning, we progressively remove the two levels of contrastive loss, thus generating three ablations, \ie \textit{w/o item}, \textit{w/o bundle}, and \textit{w/o both}. Table~\ref{tab:ablation_study} depicts the results, which demonstrate the both contrastive losses are helpful, especially on the sparser version of datasets.

\subsection{Model Study (RQ3)} \label{subsec:model_study}
To explicate more details and various properties of our method, we further conduct a list of model studies. 

\begin{table}[t]
\caption{The overall performance (NDCG@20) of our CLHE and baselines on the warm setting. "\%Improv." denotes the relative improvement over the strongest baseline.}
\vspace{-0.1in}
\label{tab:overall_performance_id}
\centering
\setlength{\tabcolsep}{1mm}{
    \resizebox{0.46\textwidth}{!}{
        \begin{tabular}{l c c c c}
        \toprule
         Models & POG & POG\_dense & Spotify & Spotify\_sparse \\
        \midrule
         \textbf{MultiDAE}       & 0.0338 & 0.2725 & 0.1527 & 0.0839  \\
         \textbf{MultiVAE}       & 0.0424 & 0.2871 & 0.0900 & 0.0384  \\
         \textbf{Bi-LSTM}        & 0.0145 & 0.1997 & 0.0946 & 0.0281  \\
         \textbf{Hypergraph}     & 0.0393 & 0.2860 & 0.0822 & 0.0401  \\
         \textbf{Tranformer}     & \underline{0.0520} & \underline{0.2969} & \underline{0.1837} & \underline{0.1199}  \\
         \textbf{TranformerCL}  & 0.0280 & 0.2747 & 0.1766 & 0.1152 \\
        \midrule
         \textbf{CLHE (ours)} & \textbf{0.0554} & \textbf{0.3218} & \textbf{0.1846} & \textbf{0.1245}  \\
         \textbf{\%Improv.}      & 6.53   & 8.38   & 0.48   & 3.81    \\
        \bottomrule
        \end{tabular}
    }
}
\vspace{-0.1in}
\end{table}

%\begin{table}
%    \centering
%    \caption{Performance (NDCG@20) analysis under the modality-missing setting.}
%    \label{tab:missing_modality_evaluation}
%    \resizebox{0.46\textwidth}{!}{
%        \begin{tabular}{lcccc} 
%        \hline
%        Models   & POG    & POG\_dense & Spotify & Spotify\_sparse  \\ 
%        \hline
%        MultiDAE    & 0.0046 & 0.1930  & 0.0836  & 0.0489  \\
%        w/o item CL & 0.0170 & 0.2449  & 0.1711  & 0.1109  \\
%        CLHE     & 0.0189 & 0.2467  & 0.1744  & 0.1163  \\
%        \hline
%        \end{tabular}
%    }
%\end{table}

\subsubsection{Cold-start Items} One of the main challenges for bundle construction is cold-start items that have never been included in previous bundles. It is difficult to directly evaluate the methods solely based on cold-start items since there are few testing bundles where both the input and result partial bundles purely consist of cold-start items. Nevertheless, we come up with an alternative way to indirectly test how these methods perform against cold-start items. Specifically, we remove all the cold-start items and just keep the warm items in the testing set, \ie the warm setting. We test our method and all the baseline models on this warm setting. The results shown in Table~\ref{tab:overall_performance_id} illustrate that: 1) the performance of all the models on the warm setting are much better than that of the warm-cold hybrid setting (the normal setting as shown in Table~\ref{tab:overall_performance_mm}), exhibiting that the existence of cold-start items significantly deteriorate the performance; and 2) the performance gap between CLHE and the strongest baseline in the hybrid setting is obviously much larger than that on the warm setting, implying that our method's strength in dealing with cold-start items.

\begin{figure}
    \centering
    \includegraphics[width = \linewidth]{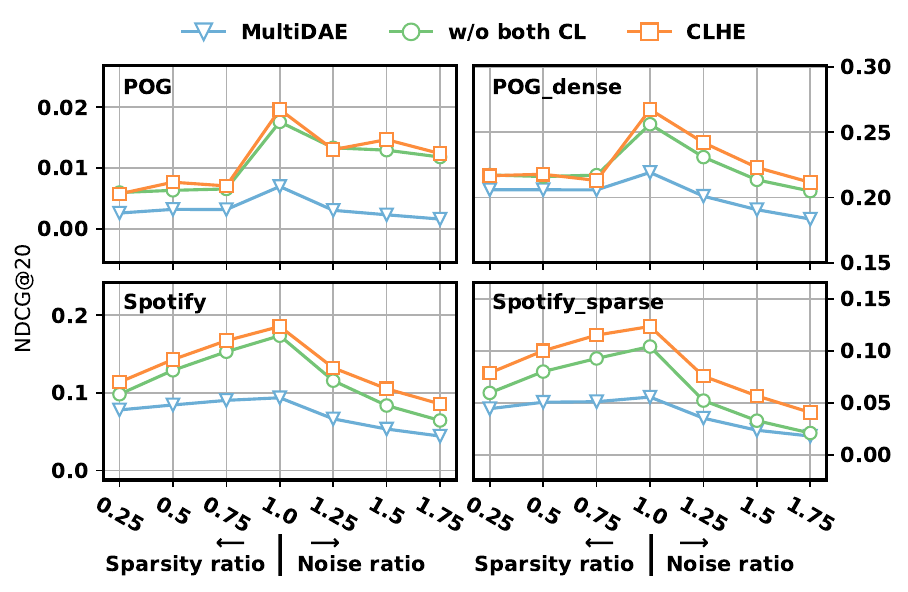}
    \vspace{-0.1in}
    \caption{Performance analysis with varying rates of sparsity and noise in the partial bundle.}
    \label{fig:noise_bundle_evaluation}
    \vspace{-0.1in}
\end{figure}

%\subsubsection{Modality Missing Items}
%The modality missing issue, that is one or more of the three types of features are unavailable, could affect the bundle construction. We argue that the item-level contrastive learning can alleviate this issue to some extent by investigating the self supervision signals. To test this property, we randomly remove some modality features 

\subsubsection{Sparsity and Noise in Bundle}
Another merit of our approach is that the contrastive learning is able to counter the sparsity and noise issue when the input partial bundle is incomplete. To elicit this property, we corrupt the testing dataset to make the input partial bundle sparse and noisy. Specifically, we randomly remove certain portion of items from the input partial bundle to make them sparser. To make the partial bundle more noisy, we randomly sample some items from the whole item set and add them to the bundle. Then we test our model and the model without both levels of contrastive loss, and the performance curves are shown in Figure~\ref{fig:noise_bundle_evaluation}, where the x-axis is the ratio of bundle size after corruption compared with the original bundle, and the ratio=1 corresponds to the original clean data. From this figure, we can derive the conclusion that: 1) with the sparsity and noise degree increasing, both our method and baselines' performance drops; 2) our method still outperforms baselines even under quite significant sparsity or noise rate, such as removing 50\% seed items or adding 50\% more noisy items; and 3) the contrastive loss in our model is able to combat the parse and noise bundle issue to some extent.

\begin{table}[t]
\begin{center}
%\vspace{-0.1in}
\caption{Study of data augmentations (NDCG@20).}
\label{tab:data_augmentation}
\vspace{-0.1in}
\resizebox{0.46\textwidth}{!}{
    \begin{tabular}{llcccc}
        \toprule
         & Setting & POG & POG\_dense & Spotify & Spotify\_sparse \\
        \midrule
        \multirow{4}{*}{Item} 
         & NA & 0.0148 & 0.2428 & 0.1712 & 0.1202 \\ 
         & FN & \underline{0.0193} & 0.2753 & 0.1735 & 0.1172 \\
         & FD & 0.0178 & 0.2763 & 0.1718 & 0.1200 \\
         & MD & 0.0162 & \underline{0.2773} & \underline{0.1806} & \underline{0.1212} \\
        \midrule
        \multirow{2}{*}{Bundle} 
         & ID & 0.0184 & \underline{0.2773} & 0.1791 & \underline{0.1212}\\ % Item Dropout
         & IR & \underline{0.0193} & 0.2750  & \underline{0.1806} & 0.1185\\ % Item Replacement
        \bottomrule
    \end{tabular}
}
\end{center}
\vspace{-0.1in}
\end{table}

\subsubsection{Data Augmentations}
Data augmentation is the crux to contrastive learning. We search over multiple different data augmentation strategies at both item- and bundle-level contrastive learning, in order to find the best-performing setting. In Table~\ref{tab:data_augmentation}, we present the performance of CLHE under various data augmentations at both item- or bundle-level. Overall speaking, data augmentation methods may affect the performance and proper selection is important for good results. 

\begin{figure}
    \centering
    \includegraphics[width = 0.99\linewidth]{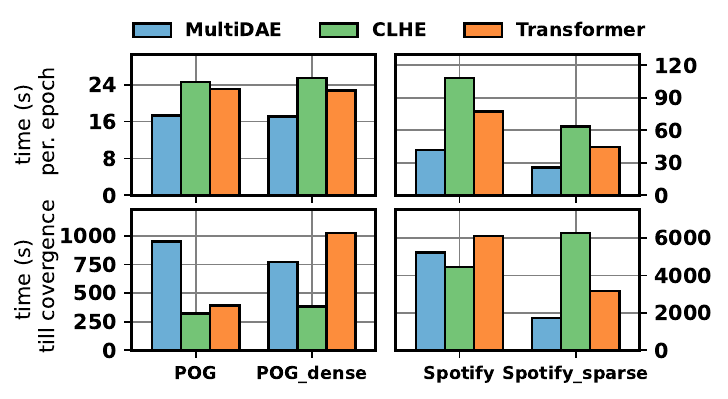}
    \vspace{-0.1in}
    \caption{Computational complexity analysis.}
    \label{fig:complexity_analysis}
    \vspace{-0.1in}
\end{figure}

\subsubsection{Computational Complexity}
Self-attention calculates every pair of instances in a set, \ie features of an item or items of a bundle, thus it usually suffers from high computational complexity. We record the time used for every training epoch and the time used from the beginning of training till convergence, and the records of our method and two baselines, \ie MultiDAE and Transformer, are illustrated in Figure~\ref{fig:complexity_analysis}. The bar chart reveals that on the one hand, our method is computationally heavy since it takes the longest time for each training epoch; on the other hand, our method takes the least training time to reach convergence on three datasets. In conclusion, our method is effective and efficient during training, while the inherent complexity induced by hierarchical self-attention may impose the inference slower. We argue that various self-attention acceleration approaches could be considered in practice, which is out of the scope of this work.

\begin{figure}
\centering
    \includegraphics[width = 0.99\linewidth]{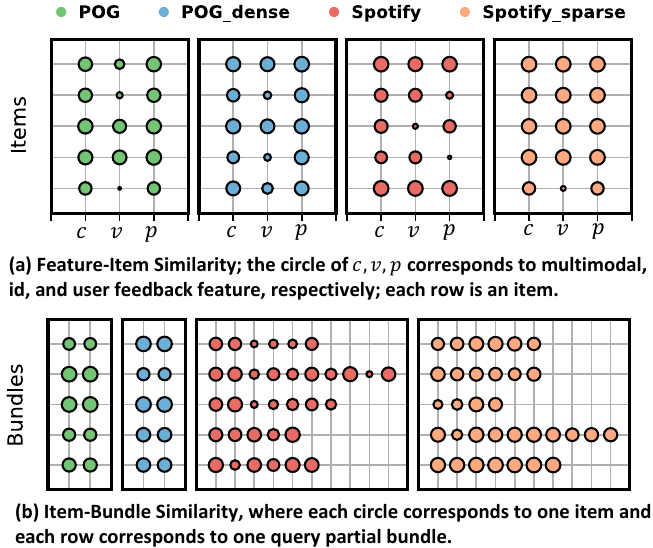}
    %\vspace{-0.1in}
    \caption{Illustration of similarity a) between each feature and the whole item representation; and b) between each item and the whole bundle representation. The size of the circles is positively correlated with its corresponding cosine similarity.}
    \label{fig:case_study}
    \vspace{-0.1in}
\end{figure}

\subsubsection{Case Study}
We would like to further illustrate some cases to portrait how the hierarchical encoder learn the association of multimodal features and the multiple items' representations. Specifically, for both item- and bundle-level self-attention modules, we take the last layer's output representation as each feature's (item's) representation, and calculate the cosine similarity score with the whole item (bundle). We cherry pick some example items and bundles, as shown in Figure~\ref{fig:case_study}. The results of feature-item similarity exhibit that the three type of features could play distinctive roles in different items, showing the importance of all the three types of features. For the item-bundle similarity results, we can find that items do not equally contribute to their affiliated bundles, thus it is crucial to model the bundle composition patterns. Here we just intuitively illustrate some hints about the bundle representation learning, more sophisticate analysis, such as feature pair or item pair co-effects, is left for future work.

\begin{figure}
    \centering
    \includegraphics[width = 0.99\linewidth]{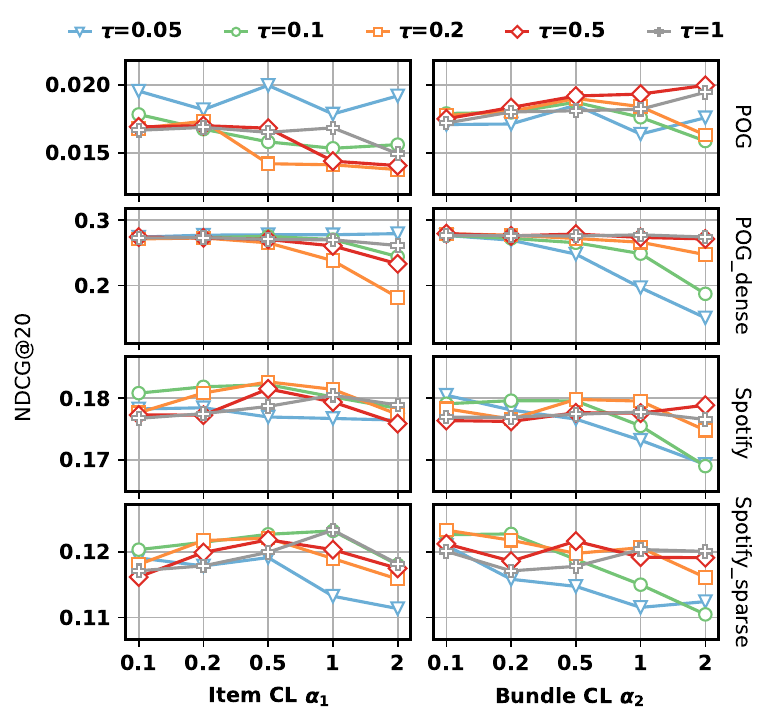}
    \vspace{-0.1in}
    \caption{Hyper-parameter analysis.}
    \label{fig:hyperparameter_analysis}
    \vspace{-0.1in}
\end{figure}

\subsubsection{Hyper-parameter Analysis}
We also present the model's performance change \wrt the key hyper-parameters, \ie the temperature $\tau$ in contrastive loss and the weights ${\alpha}_1, {\alpha}_2$ for the two contrastive losses. The curves in Figure~\ref{fig:hyperparameter_analysis} reveal that the model is still sensitive to these hyper-parameters, and proper tuning is required to achieve optimal performance.

% \begin{figure}
%     \centering
%     \includegraphics[width = 0.99\linewidth]{figures/item_viz.pdf}
%     \vspace{-0.1in}
%     \caption{Modality-level attentive feature analysis.}
%     \label{fig:item_viz}
%     \vspace{-0.1in}
% \end{figure}

% \begin{figure}
%     \centering
%     \includegraphics[width = 0.99\linewidth]{figures/bundle_viz.pdf}
%     \vspace{-0.1in}
%     \caption{Item-level attentive feature analysis.}
%     \label{fig:bundle_viz}
%     \vspace{-0.1in}
% \end{figure}
\section{Related Work} \label{sec:related_work}
We review the literature about bundles, including: 1) bundle recommendation and construction, and 2) bundle representation learning.

\subsection{Bundle Recommendation and Construction}
Product bundling is a mature marketing strategy that has been applied in various application scenarios, including fashion outfit~\cite{POG2019,HFGN2020}, e-commerce~\cite{revisitBundle2022}, online music playlist~\cite{EFM2017,SpotifyChallengeRecsys2018}, online games~\cite{BundleNet2020}, travel package~\cite{TripRec2018}, meal~\cite{MealRec2022}, and \etc. Personalized bundle recommendation~\cite{DAM2019,BGCN2021,CrossCBR2022} is the pioneering work that first focuses on bundle-oriented problems in the data science community. 
%Specifically, people are interested in recommending the predefined bundles to users, which is an extension of recommendation from individual items to bundles. 
Soon after that, researchers realize that just picking from predefined bundles cannot satisfy people's diverse and personalized needs. Thereby, the task of personalized bundle generation~\cite{BGN2019,BGGN2023,BYOB2021,Conna2022,CAR2020,TOG2023} is naturally proposed where the model aims to automatically generate a bundle from a large set of items catering to a given user. It has to simultaneously deal with both users' personalization and item-item compatibility patterns, where the user-item interaction is specifically utilized for personalization modeling. In this paper, we only focus on bundle construction, which is committed to generate more bundles to enrich the bundle catalog for the platform. In addition, most of the bundle-oriented research in general domain still falls into the id-based paradigm, where very few domains, such as the fashion domain, have explored multimodality. We extend the multimodal learning to one more domain of music playlist.
%thus emphasizing that it is crucial to incorporate multimodal information into general domains. 
Moreover, we also leverage user feedback to multimodal bundle construction.
%, which would setup a new but more comprehensive formulation for bundle construction.

\subsection{Bundle Representation Learning}
Bundle representation learning is the crux of all the bundle-oriented problems. 
%Essentially, bundle representation learning is a composition optimization problem, where the model is required to generate a good representation considering all the possible correlations among all the items within a bundle. 
Initial studies~\cite{FPMC2010} treat a bundle as a special type of item and just use the bundle id to represent it. Naturally and reasonably, people get to consider the encapsulated items within a bundle to generate more detailed representation. The simplest method is performing average pooling over the included items~\cite{MultiDAE2016}. Later on, sequential models, such as Bi-LSTM~\cite{Bi-LSTM2017}, are utilized to capture the relations between two consecutive items. However, the items within a bundle are not ordered essentially, and sequential models cannot well capture all the pair-wise correlations. To address the limitation, attention models~\cite{DAM2019,OutfitNet2020,CAR2020}, Transformer~\cite{PMGT2021,PEAR2022,BRUCE2022,BundleGT2023,OutfitTransformer2023,Transformer2017} and graph neural networks (GNNs)~\cite{BGCN2021,DGMAE2023,MIDGN2022,Hypergraph2022,DGSR2021} are leveraged to model not only every pair of items within a bundle, but also the higher-order relations by stacking multiple layers. 
%In this paper, we follow the established model of self-attention to model the bundle. 

Even though many efforts have been paid to the item correlation learning to achieve good bundle representation, the multimodal information has been less explored. Multimodal information, such as textual, visual, or knowledge graph information of items, demonstrates to be effective in general recommendation~\cite{MMGCN2019,wei2023lightgt,wei2022causal}. In the fashion domain, visual and textual features have been extensively investigated for pairwise mix-and-match~\cite{TransNFCM,MMOCM2021} or outfit compatibility modeling~\cite{NGNN2019,MOCM-MGL2023}. However, these works have not been extended to other domains, such as music playlist, where the audio modality has been rarely studied in the bundle recommendation or construction problem. More importantly, we argue that the user-item interaction information, which is widely utilized in the personalized recommendation problem, can serve as an additional modality in bundle construction. Sun \etal~\cite{revisitBundle2022} leverage a pre-trained CF model to obtain item representation to enhance the bundle completion task, while they have not fully justify the rationale and motivation. 
To the best of our knowledge, none of the previous works put together all the user-item interaction, bundle-item affiliation, and item content information for bundle construction. 

%Recently, contrastive learning has been leveraged in various self-supervised learning tasks, including computer vision~\cite{SimCLR2020}, natural language processing~\cite{SimCSE2021}, recommender system~\cite{SGL2021}, and bundle-oriented works~\cite{MIDGN2022,CrossCBR2022}. We inherit this general powerful capability to enhance the representations of both item and bundle.
\section{Conclusion and Future Work} \label{conclusion}

In this work, we systematically study the problem of bundle construction and define a more comprehensive formulation by considering all the three types of data, \ie multimodal features, item-level user feedback data, and existing bundles. Based on this formulation, we highlight two challenges: 1) how to learn expressive bundle representations given multiple features; and 2) how to counter the modality missing, noise, and sparity problem. To tackle these challenges, we propose a novel method of \textbf{C}ontrastive \textbf{L}earning-enhanced \textbf{H}ierarchical \textbf{E}ncoder (CLHE) for bundle construction. Our method beats a list of leading methods on four datasets of two application domains. Extensive ablation and model studies justify the effectiveness of the key modules.

Despite the great performance that has been achieved by this work, there is still large space to be explored for bundle construction. First, the current evaluation setting is a little bit rigid and inflexible, it is interesting to extend it to more flexible setting to align with real applications. For example, given arbitrary number of seed items, the model is asked to construct the bundle. Second, some of the feature extractors are pre-trained and fixed, \ie the multimodal feature extraction and user-item interaction models. Is it possible to optimize these feature extractors in an end-to-end fashion thus the extracted features would be more aligned to the bundle construction task? Finally, this work just targets at unpersonalized bundle construction. It is an interesting and natural direction to push forward this work to personalized bundle construction.
\section*{acknowledgement}
This research is supported by NExT Research Center, National Natural Science Foundation of China (9227010114), and the University Synergy Innovation Program of Anhui Province (GXXT-2022-040).

%\newpage
\bibliographystyle{ACM-Reference-Format}
%\balance
\bibliography{0_main}
\end{document}